\documentclass[pre,aps,preprintnumbers,amsmath,amssymb,superscriptaddress,twocolumn]{revtex4-2}
\usepackage{graphicx,subfigure}
\usepackage{pstricks}
\usepackage{xcolor}
\usepackage{enumerate}
\usepackage{ulem}
\usepackage[mathlines]{lineno}
\begin{document}

\title{Polymerization and replication of primordial RNA induced by clay-water interface dynamics}

\author{Carla Alejandre}
\affiliation{These authors contributed equally}
\affiliation{Centro de Astrobiolog\'{\i}a (CAB), CSIC-INTA, Carretera de Ajalvir km 4, 28850 Torrej\'on de Ardoz, Madrid, Spain}
\affiliation{Grupo Interdisciplinar de Sistemas Complejos (GISC), Madrid, Spain}
\author{Adri\'an Aguirre-Tamaral}
\affiliation{These authors contributed equally}
\affiliation{Centro de Astrobiolog\'{\i}a (CAB), CSIC-INTA, Carretera de Ajalvir km 4, 28850 Torrej\'on de Ardoz, Madrid, Spain}
\affiliation{Department of Biology, University of Graz, Universitätsplatz 2, 8010 Graz, Austria}
\author{Carlos Briones}
\affiliation{Centro de Astrobiolog\'{\i}a (CAB), CSIC-INTA, Carretera de Ajalvir km 4, 28850 Torrej\'on de Ardoz, Madrid, Spain}
\author{Jacobo Aguirre}
\affiliation{Centro de Astrobiolog\'{\i}a (CAB), CSIC-INTA, Carretera de Ajalvir km 4, 28850 Torrej\'on de Ardoz, Madrid, Spain}
\affiliation{Grupo Interdisciplinar de Sistemas Complejos (GISC), Madrid, Spain}
\affiliation{To whom correspondence should be addressed. E-mail: jaguirre@cab.inta-csic.es}

\begin{abstract}
In the study of life's origins, a key challenge is understanding how RNA could have polymerized and subsequently replicated in early Earth.
We present a theoretical and computational framework to model the non-enzymatic polymerization of ribonucleotides and the template-dependent replication of primordial RNA molecules, at the interfaces between the aqueous solution and a clay mineral. 
Our results demonstrate that systematic polymerization and replication of single-stranded RNA polymers, sufficiently long to fold and acquire basic functions ($>$15 nt), were feasible under these conditions.
Crucially, this process required a physico-chemical environment characterized by large-amplitude oscillations with periodicity compatible with spring tide dynamics, suggesting that large moons may have played a role in the emergence of RNA-based life on planetary bodies. 
Interestingly, the theoretical analysis presents rigorous evidence that RNA replication efficiency increases in oscillating environments compared to constant ones. Moreover, the versatility of our framework enables comparisons between different genetic alphabets, showing that a four-letter alphabet ---particularly when allowing non-canonical base pairs, as in current RNA--- represents an optimal balance of replication speed and sequence diversity in the pathway to life. 
\end{abstract}

\maketitle

\section{Introduction} \label{sec:intro}
The origin of life on Earth is a fundamental question that remains elusive when we aim to fully describe it in physico-chemical terms. 
For a complex enough molecular system to be considered alive, it must exhibit and combine a number of features such as compartmentalization, energy dissipation, replication of a genetic material and metabolism, thus allowing its self-reproduction and open-ended evolution. 
In the earlier stages of the precellular world, simple mechanisms at the molecular level, still not catalyzed by enzymes, must have driven the increasing complexity of prebiotic molecules. Different sets of chemical species needed to be concentrated and separated from the external environment through some form of surface- or membrane-based compartmentalization. A small fraction of them could have been used as monomers to build (by means of non-enzymatic polymerization mechanisms) progressively longer and more complex polymeric molecules, some of which were endowed with heritable information that provided a substrate for Darwinian evolution to act upon~\citep{ruiz2014prebiotic,sutherland2016origin,kitadai2018origins,preiner2020future}. 

Since the pioneering work of Aleksandr I. Oparin a century ago~\citep{ai1924proiskhozhdenie} and of John B.S. Haldane five years later~\citep{haldane1929origin}, particularly over the past four decades, various hypotheses have emerged to try to understand the transition from chemistry to biology. One model that gained significant attention since the 1980s is the RNA World hypothesis~\citep{gilbert1986origin, robertson2012origins,pearce2017origin,Cech2012,Bose2022}. According to its current version, in the framework of prebiotic systems chemistry, the first self-reproducing entities contained single-stranded RNA (ssRNA) as genetic material, as well as a proto-metabolism based on ribozymes: structured ssRNA molecules endowed with catalytic activities (including RNA ligase and RNA polymerase), likely assisted by short abiotic peptides and low molecular weight compounds as cofactors~\citep{szostak2011optimal,ruiz2014prebiotic,joyce2018protocells}. 
 
However, before the establishment of an RNA (or RNA-peptide) world, the available and, somehow, chemically activated ribonucleotides (from now on, termed nts) should have polymerized into ssRNA oligomers in absence of the catalytic activity of ribozymes. Additionally, a non-enzymatic process of template-dependent RNA replication could have ensured the availability of populations of RNA molecules with related sequences, on which selection could act~\citep{leslie2004prebiotic}.
Indeed, the minimum length of an RNA polymerase ribozyme (i.e., a structured RNA molecule capable of replicating other RNA molecules used as templates) has been estimated in the range of 165 nts~\citep{johnston2001rna} to 182 nts~\citep{portillo2021witnessing, papastavrou2024rna, chen2024rna},
which is far longer than the up to 50-mer ssRNA molecules that can be experimentally obtained by means of non-enzymatic polymerization using activated nts and montmorillonite clay as a catalyst~\citep{huang2003synthesis, huang2006one,Kloprogge2022}.
Therefore, random RNA polymerization of short oligomers, which is known to be favored by clays and other heterogeneous media~\cite{Ferris2006, Jerome2022, Himbert2016, Attwater2010}, could have been followed by a ligation-based, stepwise process of modular evolution of RNA that allowed the assembly of a template-dependent RNA polymerase ribozyme~\cite{Briones2009, wachowius2019non}. 
Importantly, non-enzymatic, template-dependent RNA replication has been explored both experimentally~\cite{Deck2011, Adamala2013, Adamala2015, OFlaherty2019} and through theoretical models, most of them focusing on liquid environments~\cite{Anderson1983, Rokhsar1986, Orgel1992, Fernando2007, Tkachenko2015, Fellermann2017,Toyabe2019, Tupper2021, Rosenberger2021, Zhou2021, Juritz2022, Chamanian2022, Goppel2022}. However, little work has focused on the entire process, connecting RNA polymerization and template-dependent replication in the same heterogeneous environment and making use of a global, systems chemistry approach~\cite{Walker2012,Kaddour2014,ruiz2014prebiotic}. Furthermore, the holistic perspective of complexity science is just starting to be exploited in this context, and our work is a step in this promising direction~\cite{Garcia_2022,Bianconi:2023}.

The vast diversity of geological environments that coexisted in early Earth opens up a multitude of possible chemical reactions and physical scenarios, not all equally conducive to the establishment of an RNA world. 
Under specific conditions, ssRNA polymers can grow when wet/dry cycles are present ~\cite{morasch2014dry, Vsponer2021} or in environments subjected to extreme temperature changes ~\cite{mast2013escalation, kreysing2015heat}.
 Interestingly, extensive work suggests that oscillating environments were crucial for the emergence of life~\cite{lahav1978peptide,Kompanichenko2012,Damer2015,kaddour2018nonenzymatic,senatore2022modelling, ianeselli2023physical}.
These fluctuating conditions were common in early Earth and could be of various types. For example, analyses of tidal interactions in the Earth–Moon system since its formation suggest that, when life first emerged on our planet, a day lasted approximately 10–12 hours, resulting in tidal maxima at ocean shores every 5–6 hours~\cite{Farhat:2022}. While Lathe proposed two decades ago that tides could have provided the oscillating driving force for the origin of replicating biopolymers~\cite{lathe2004fast, lathe2005tidal}, Fernando {\it et al.} and Chamanian {\it et al.} noted that linear templates subjected to external oscillations might provoke a biopolymer elongation that would hinder its accurate replication~\citep{Fernando2007, Chamanian2022}. 
Anyway, temperature oscillations and wet/dry cycles have received the most attention: temperature cycles are known to promote the hybridization/denaturation of complementary ssRNA strands~\cite{Szostak2012}, while dehydration periods are essential to allow condensation reactions, such as the formation of $3'$-$5'$ phosphodiester bonds necessary for RNA polymerization~\cite{Damer2015,Higgs2016,ianeselli2022water}. 

Numerous studies have experimentally demonstrated that the surfaces of certain minerals and rocks, in particular clays such as montmorillonite, exhibit chemical and crystallographical properties (including their microscopic structure, distribution of cations and spacing of the phyllosilicate interlayers) that can favor the adsorption and correct orientation of nts while providing a locally anhydrous environment, thus catalyzing the abiotic polymerization of RNA~\cite{ferris1996synthesis,huang2003synthesis, huang2006one, Ferris2006, aldersley2011role, jelavic2017prebiotic, kaddour2018nonenzymatic, Jerome2022,Kloprogge2022}. 
The importance of including clay surfaces in the prebiotic context is further highlighted by studies showing that they
protect molecules from degradation~\cite{ruiz2014prebiotic} and concentrate organic molecules~\cite{Bernal1949}. Also, clays catalyze different key processes, including 
the synthesis
of nucleobases from their precursors~\cite{ruiz2014prebiotic}, the formation of vesicles from fatty acid micelles~\cite{Hanczyc2003}, and the polymerization of amino acids into peptides~\cite{lahav1978peptide}. Furthermore, the sedimentation of ssRNA oligomers on clay surfaces or other minerals induces a phase separation into oligonucleotide-dense and dilute phases that can facilitate the selection of specific RNA sequences with prebiotic functions~\cite{bartolucci2023sequence}. 

Despite the aforementioned research conducted on the comparison among potential physico-chemical contexts for the origin of life, we are still far from unveiling the precise mechanisms that led to the emergence and replication of the first RNA polymers, around 4 Gyr ago~\cite{Moody:2024,kitadai2018origins,pearce2018constraining}. To shed light on this subject, we propose an integrative numerical and theoretical approach that considers both non-enzymatic RNA polymerization and template-dependent replication in an environment where nts and ssRNA polymers interact through a clay-water interface.
We introduce EarlyWorld, a computational framework supported by a theoretical model, which allows us to test how a wide variety of environmental conditions could affect the polymerization and replication of ssRNA oligomers. This approach differs from most existing models of non-enzymatic RNA replication in fluctuating conditions, which take the existence of two phases for granted: a ``cold phase'' where elongation reactions take place, and another ``warm phase'' where all RNA duplexes denature. EarlyWorld has been developed as a toy-model representation of the chemical processes that could have occurred in a clay-water microenvironment in early Earth, and it is not pretended to be chemically detailed at this stage, but to capture the essential features of the processes modeled.  
By exploring and refining our understanding of the specific fluctuating environments that could effectively promote the increase in biochemical complexity, we aim to certify whether an heterogeneous mineral-water interface in early Earth was suitable for the emergence of RNA populations, as a prerequisite for the origin of life.

\section{Methods: Description and rules of the model}
\begin{figure*}[ht!]
\centering
\includegraphics[width=\linewidth]{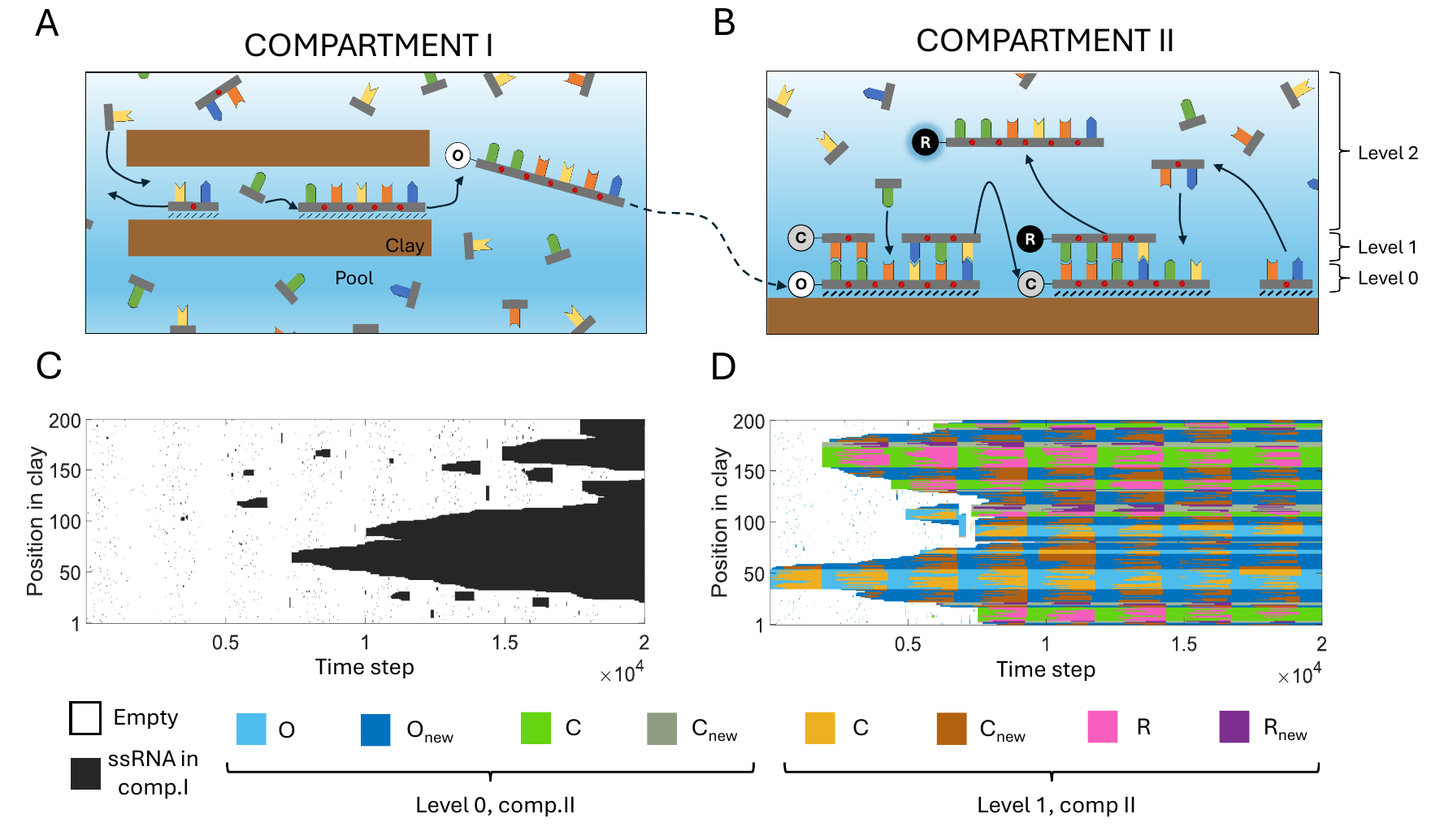}
\caption{Description of the non-enzymatic RNA polymerization and replication in an early Earth's clay-water interface modeled by EarlyWorld.
Top panels (A and B) sketch each natural compartment of the model and bottom panels (C and D) show example realizations of the evolution with time $t$ of all the positions available in an oscillating clay environment. The bottom legend corresponds to the colors shown in panels C and D. If not said otherwise, throughout this work we will use clays of length $L=200$ positions, $N=1600$ nts equally distributed between A, G, C, U types and initially placed in the aqueous phase, C$\equiv$G and A$=$U nucleobase complementarities, and simulation times $t_\mathrm{max}=2\times10^4$ time steps. 
Compartment I: Random polymerization occurs due to the catalytic action of clays in an interlayer of a clay grain, where complementary base pairing is not permitted due to steric hindrance. Only clay-nt interactions (of strength $\alpha$) are permitted and ssRNA can be formed at the clay surface. Compartment II: The template replication of the original strand O occurs involving complementary base pairing; strand C (complementary to O) is formed, denatured from the template, released to the pool and adsorbed to the clay. In a next step, the replication of strand C yields strand R, whose sequence is equal to the original (R = O). Once the dsRNA molecule C-R is formed, its denaturation will release the ssRNA oligomer R, completing the RNA replication process. Other polymers O$_\mathrm{new}$ spuriously formed in comp.II can also be used as template (C$_\mathrm{new}$) and replicated (R$_\mathrm{new}$). In comp.II both clay-nt (of strength $\alpha$) and nt-nt (of strength $\beta$) interactions are allowed, and ssRNA molecules can remain adsorbed to the clay (level 0), hybridize to their complementary strand to form dsRNA (level 1), or move to the aqueous phase (level 2). In this example, $\alpha=0.8 + 0.5\sin{(2\pi t/2500)}$ and $\beta=6 + 5.9\sin{(2\pi t/2500)}$. 
}
\label{fig1}
\end{figure*}

We have developed a computational environment called EarlyWorld to simulate the non-enzymatic RNA polymerization and template replication in primordial heterogeneous media, such as on the interface between an aqueous solution and a clay surface, in the absence of RNA polymerase ribozymes. 
Clays are known to present a complex structure of internal channels and interlayers, where water can get in and promote an intricate set of clay-water interfaces of around 100 m$^2$ per g of clay~\cite{macht2011specific}.
The RNA polymerization and replication processes computationally simulated here take place in two compartments of the clay showing different geochemical properties.
Figure~\ref{fig1} schematizes the dynamics described by EarlyWorld: Fig.~\ref{fig1}A and B sketch the natural compartments of the model, and Fig.~\ref{fig1}C and D show example realizations of the evolution with time of the system in each compartment. A rigorous explanation of the algorithm is provided in Supplementary Note S1, and information on the public availability of the code is found in section Code Availability.

\subsection{Compartment I}
Random polymerization of nts takes place in compartment I (comp.I from now on) due to the catalytic action of clays. 
It represents an internal channel or interlayer that is wide enough to allow ssRNA, but not double-stranded RNA (dsRNA) oligomers to pass through. This assumption is supported by experimental data showing that the A-helix of dsRNA has a diameter of $\sim2.4$ nm~\cite{lipfert2014double}, while the interlayer space of montmorillonite and other clays typically varies between 0.9-1.0 nm (in the dried state) to 1.3 nm (hydrated with one water layer) and 1.9 nm (fully hydrated with 2-3 water layers), which also depends on their ionic composition, pH and presence of different cations in the medium~\cite{ervithayasuporn2019modifying,Brigatti2013}.
In consequence, complementary base pairing (and, thus, template-dependent replication) is not possible in the interlayers due to steric hindrance (Fig.~\ref{fig1}A). We have focused on interlayers of this kind as the preferred location for ssRNA polymerization, following the assumption that the potential presence of catalytic sites in the montmorillonite interlayers enhances the rate and regioselectivity of polymerization~\cite{Ferris2006,mathew2010influence,DeOliveira2021}, though alternative scenarios are also considered (see Discussion). 

The compartment has two different levels: the clay-water interface (shortened to ``clay''), and the aqueous phase (or ``pool''). For simplicity, the clay level is described by a 1D-vector of length $L$ where each element/position can be either empty or contain a single nt, while the aqueous phase is a {\it disordered box} with space for an unlimited number of nts and ssRNA molecules.
The clay and the pool will be enriched with ssRNA oligomers throughout each simulation, but every numerical realization in comp.I starts with the clay level totally empty and a number $N$ of nts equally distributed between those composing the genetic alphabet in use (EarlyWorld accepts any genetic alphabet and base pairs of all kind, see section Results~\ref{sec:alphabets}). 
At each time step, one of the molecules in the pool is chosen randomly and gets adsorbed to the clay in a randomly chosen place (involving one or more contiguous positions), as far as it is empty. Otherwise the molecule returns to the pool. In the clay, it will form a covalent ($3'$-$5'$ phosphodiester) bond with other nts potentially placed at any (or both) of its adjacent positions. The resulting polymer remains attached to the clay, and we assume that its covalent sugar-phosphate backbone 
is unbreakable (except when we explicitly study the influence of hydrolysis in the system, see section Results~\ref{sec:compII}). After that, every oligomer adsorbed to the clay can be released to the pool with a desorption probability of
\begin{equation}
\centering
P_{\alpha}=e^{-l\alpha},
\label{Prup1}
\end{equation}
where the interaction parameter $\alpha>0$ describes the strength of clay-nt 
interactions (due to Van der Waals forces and ionic bonds) in the environment that surrounds them, and $l$ is the length (i.e., number of nts) of the polymer whose desorption is evaluated. 
Finally, note that only the oligomers detached from the clay into the pool are recorded as effectively produced in the process, because exclusively free and mobile ssRNA molecules can migrate to compartment II and continue the process. 

\subsection{Compartment II} This compartment represents a wide internal channel, an inter-particle site or an external clay surface, and, in consequence, is able to accommodate both ssRNA and dsRNA molecules and template-dependent RNA replication based on complementary base pairing.
Simulations start with the same number of nts of each type available in the pool and with an initial ssRNA molecule (original strand O from now on), consisting of 20 randomly chosen nts adsorbed to the clay in a random place.
This ssRNA O represents a strand that is assumed to be formerly polymerized in comp.I, was released from it, eventually entered compartment II (comp.II from now on) and attached to the clay surface. Three levels are relevant for the evolution of the system: the clay or ``level 0'', where nts and polymers can be adsorbed (as in comp.I); the complementary level or ``level 1'', where one nt can only be placed if it is complementary to other nt included in a ssRNA oligomer which is placed in level 0 and thus contribute to the formation of dsRNA via hydrogen bonding; and the aqueous phase, pool or ``level 2'', where free nts, ssRNA and dsRNA stay in solution.
In each time step, a randomly chosen molecule in the pool interacts with a randomly chosen place in the surface of the clay. ssRNA and dsRNA of any length can be adsorbed to the clay in level 0, and ssRNA will be hybridized to other ssRNA strand adsorbed to the clay acting as a template in level 1 if they are fully complementary. After that, every ssRNA or dsRNA of length $l$ in level 0 will have a desorption probability (i.e., probability of being released to the pool) of $P_{\alpha}=e^{-l\alpha},$ where $\alpha$ is the clay-nt interaction parameter already mentioned in comp.I. Also, every dsRNA (both attached to the clay or free in the pool) can be denatured with a probability of
\begin{equation}
\centering
P_{\beta}=e^{-l\beta},
\label{Prup2}
\end{equation}
where $l$ stands for the dsRNA length (i.e., number of nts), and the nt-nt interaction parameter $\beta>0$ describes the strength of the hydrogen bond interactions established between their nucleobases in the environment that surrounds them.

Throughout each simulation, in comp.II the dynamics described might eventually release complementary C strands of the original O sequence to the pool (level 2), as well as complementary C$_\mathrm{new}$ strands of other newly-generated O$_\mathrm{new}$ sequences randomly polymerized in the clay (level 0). Interestingly, when a polymer C falls in an empty place of the clay and is adsorbed, it might become the template for a further polymerization based on nucleobase complementarity, and the denaturation of the dsRNA formed will release to the pool a ssRNA that is ``complementary to the complementary C'': the strand replicate or R, equal (partially, or in its full length) to the original polymer O.

In summary, our model describes the influence of the environment in the polymerization and replication of short ssRNA polymers through two parameters, $\alpha$ and $\beta$, which represent the global strength of clay-nt and nt-nt interactions, respectively. Changing these parameters mimics varying external physico-chemical magnitudes such as temperature, pH, ionic strength and presence of divalent cations, to cite a few. Large values of $\alpha$ and $\beta$ represent mild environments that facilitate said interactions (and the concomitant formation of progressively longer ssRNA and dsRNA molecules), whereas low values represent harsh environments and have the opposite effect. 

\section{Results}
\subsection{Compartment I: RNA polymerization on a clay surface}
\begin{figure}[ht]
\centering
\includegraphics[width=\linewidth]{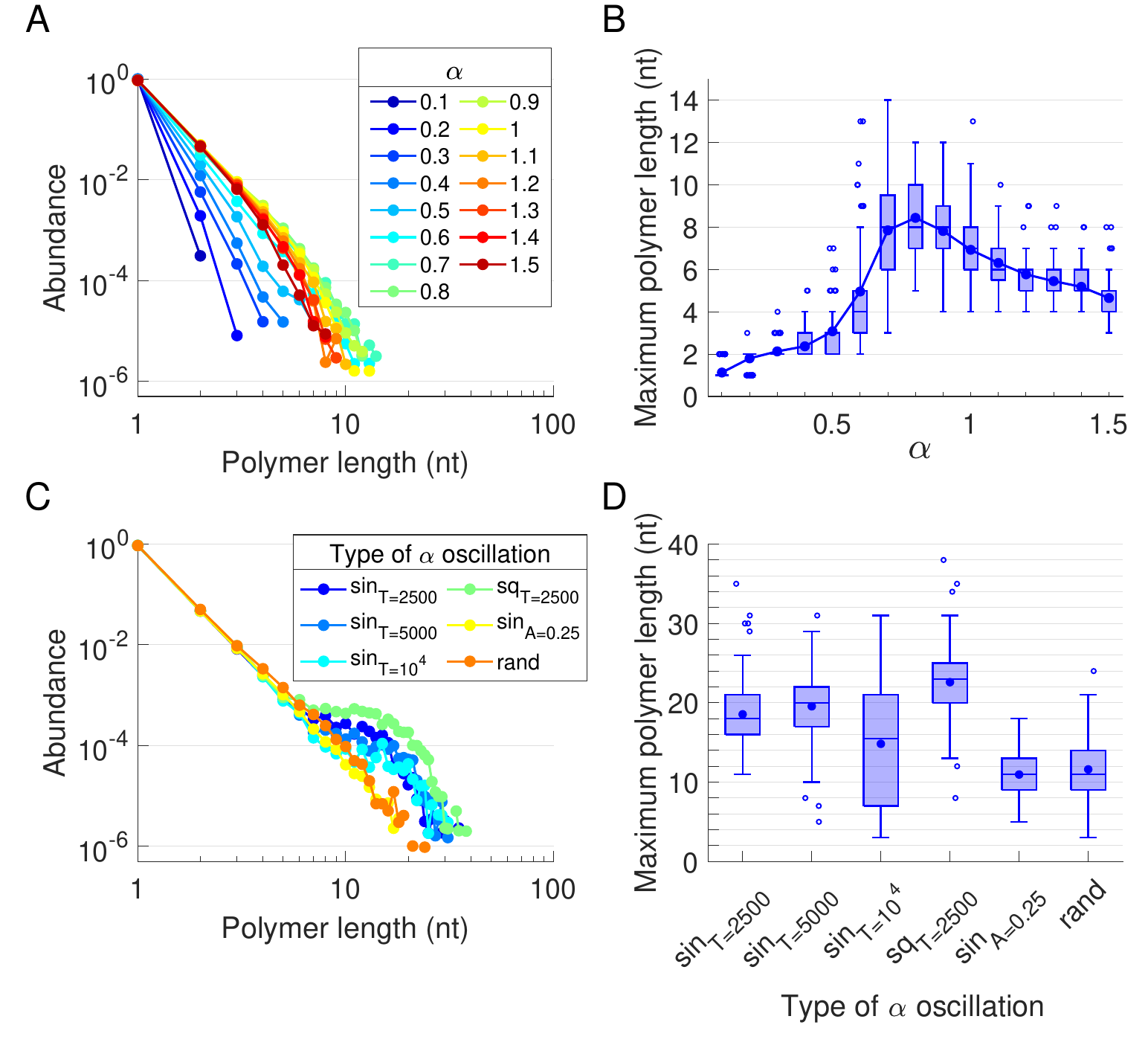}
\caption{Effect of constant and fluctuating environmental conditions on the polymerization of RNA catalyzed by a clay surface (comp.I). (A) Distribution of abundances of ssRNA k-mers (monomers, dimers, etc) and (B) maximum polymer length distributions over constant environments (clay-nt parameter $\alpha$ is constant with time). (C) Distribution of abundances of k-mers and (D) maximum polymer length distributions over six fluctuating environments ($\alpha$ varies with time):  ``sin'' stands for sinusoidal oscillation, ``sq'' for square wave oscillation and ``rand'' for random fluctuation. If not said otherwise in the axes, $\alpha_0=0.8$, $A=0.5$, $T=2500$. Every simulation was repeated 100 times.
}
\label{fig2}
\end{figure}

The polymerization process in comp.I depends exclusively on the clay-nt parameter $\alpha$. Figure~\ref{fig2} plots the length distribution of the ssRNA molecules (polymers from now on) that can be formed with the rules set of EarlyWorld for different environmental conditions, both constant and fluctuating. 
The length distribution of the polymers produced attached to the clay and
desorbed back to the pool at any time step for different constant values of $\alpha$ reveals a steep decay in Fig.~\ref{fig2}A, in full agreement with the assumed relationship between the abundance of polymerized ssRNA molecules and their length~\cite{Rosenberger2021, Zhou2021, Flory1953}. In this work we are specially interested in obtaining the largest polymer that can be produced in a given heterogeneous environment, as it could eventually be more structurally complex and functionally relevant than the shorter ones. Thus, Fig.~\ref{fig2}B shows the distribution of the maximum polymer length obtained along
every realization for each parameter $\alpha$ (the detailed explanation of how maximum polymer lengths are obtained is provided in Supplementary Note S2).
For $\alpha\rightarrow0$ the environment is so harsh that every nt that reaches the clay is immediately desorbed from it and released back to the pool, while for $\alpha\rightarrow\infty$ the mild environment allows the adsorbed nts to polymerize fast and become strongly stuck to the clay surface. Both limits result in the presence of very short ssRNA oligomers in the pool. For intermediate $\alpha$ values, however, oligomers can be substantially polymerized but still be desorbed, reaching maximum polymer lengths of $\sim5-12$ for $\alpha\sim 0.8$.

Fluctuating environments can be easily introduced in EarlyWorld by changing the value of the parameters $\alpha$ and $\beta$ over time. Figure~\ref{fig2}C and D show,
respectively, the length distribution of ssRNA molecules polymerized and desorbed from the clay, and the distribution of maximum polymer lengths obtained in every realization, for several $\alpha$-oscillating environments. 
Considering environmental oscillations of period $T$ and amplitude $A$, the sinusoidal oscillatory environments studied follow $\alpha(t)=\alpha_0+A \sin{(2\pi t / T)}$, square wave oscillations follow $\alpha(t)= \alpha_0+A\operatorname {sgn} \left(\sin {(2\pi t / T)}\right)$,
and randomly fluctuating environments follow the continuous uniform distribution $\alpha(t)=\displaystyle U\{\alpha_0-A,\alpha_0+A \}.$
The length distribution of polymers plotted in Fig.~\ref{fig2}C shows a power-law decay for the random environment that behaves as a lower limit for the rest of fluctuating environments, which, depending on their characteristics, show a larger or smaller hump for long ($>10$ nt) polymer lengths. 
In Fig.~\ref{fig2}D, it can be seen that, for all oscillating patterns, maximum polymer lengths notably increase with respect to those achieved in constant environments: means move from $1-9$ nts for constant environments (with absolute maxima around 15 nts) to $10-24$ nts for fluctuating ones (with absolute maxima around 40 nts).

\subsection{Compartment II: Template-dependent RNA replication on a clay surface}
\label{sec:compII}
The simplest secondary structures that ssRNA molecules can fold into are called stem-loops, consisting of a dsRNA stem closed by a terminal loop. In turn, hairpin structures are slightly more complex and contain a bulge within the stem, which can provide an additional biochemical functionality to the molecule~\cite{Svoboda2006}. Thus, it is commonly assumed that ribozymes should contain at least $15-35$ nts to be able to fold into complex enough structures~\cite{Briones2009}. As shown in the previous section, this polymer length interval can be typically obtained in EarlyWorld when the environment (parameter $\alpha$) fluctuates. Accordingly, simulations in comp.II start with a 20 nt-long, random ssRNA polymer O adsorbed to the clay. Along each simulation, we record the complementary C and replicate R strands derived from the original O sequence, which are produced and released to the pool. 

Throughout this work, two terms will be used to quantify the goodness of the overall template-dependent RNA replication process. The first one is the {\it efficiency} of the replication associated with each combination of interaction parameters $\alpha$ and $\beta$. It will be quantified through the maximum polymer length distribution obtained for these parameters. Accordingly, an environment will be said to permit efficient replication if it yields maximum polymer length distributions of large mean with several realizations reaching length 20. The second relevant term is the {\it accuracy} required to accept a certain copy as sufficiently pure. Remarkably, Cs and Rs can contain, in addition to the exact copy of the corresponding template, other random or spurious sequences attached to their two ends during the replication process, thus forming a longer molecule than the polymer O from which they derive~\cite{Fernando2007}. For that reason, and in order to avoid the loss of identity of the RNA molecules that are being replicated, throughout this work we consider those C and R polymers that have an {\it accuracy} of $>$75\%, meaning that their sequence is formed by $>$75\% of products of the original polymer O (in other words, these copies can have as much as 25\% of their length 'contaminated' by spurious sequences at their ends, see Supplementary Note S3 and Fig.~S3 for detailed information on sequence accuracy and simulations with 100\% accuracy, which yield equivalent results). 

\begin{figure}[ht!]
\centering
\includegraphics[width=\linewidth]{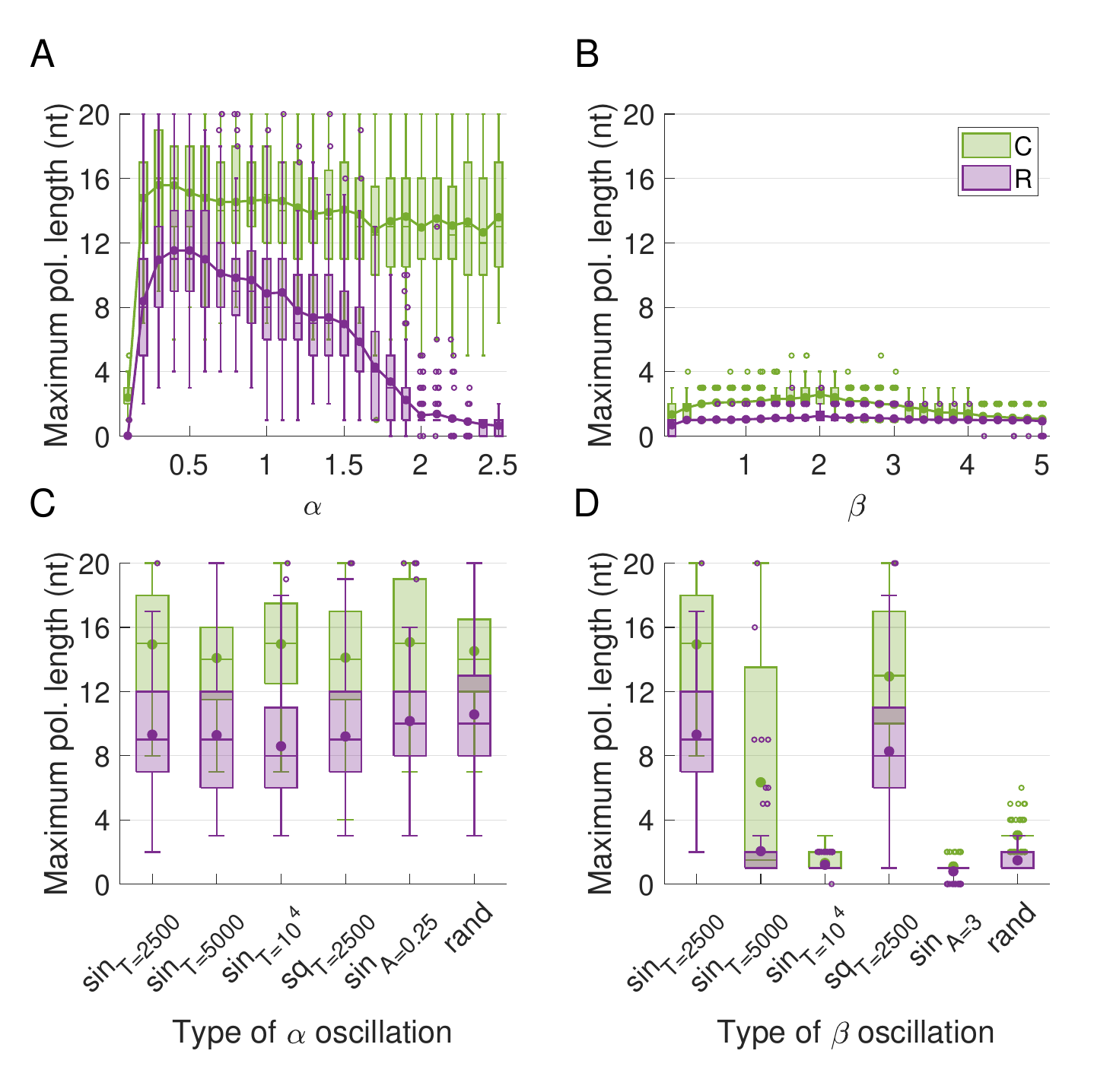}
\caption{Effect of environmental fluctuations on the template-dependent replication of a ssRNA molecule adsorbed to a clay surface (comp.II). (A) Maximum complementary (C, green in all panels) and replicate (R, purple in all panels) polymer length distributions for an oscillating $\beta$ and a constant $\alpha$ parameter. (B) Maximum polymer length distributions for an oscillating $\alpha$ and a constant $\beta$ parameter. (C) Maximum polymer length distributions for different $\alpha$ fluctuating patterns, with a fixed $\beta$ oscillation. (D) Maximum polymer length distributions for different $\beta$ fluctuating patterns, with a fixed $\alpha$ oscillation. Fluctuating environments: ``sin'' stands for sinusoidal oscillation, ``sq'' for square wave, ``rand'' for random fluctuation. If not said otherwise, $\alpha_0=0.8$, $A_{\alpha}=0.5$, $T_{\alpha}=2500$ and $\beta_0=6$, $A_{\beta}=5.9$, $T_{\beta}=2500$. Every simulation started with a random 20 nt-long ssRNA molecule, and was repeated 100 times. }
\label{fig3}
\end{figure}

Both clay-nt parameter $\alpha$ and nt-nt parameter $\beta$ play a role in the formation of C and R polymers in comp.II. Figure~\ref{fig3} shows the effect of fluctuating environmental conditions (i.e., $\alpha$ and $\beta$) on the stepwise replication of the original 20 nt-long polymer O. Considering oscillations of period $T_\gamma$ and amplitude $A_\gamma$, the sinusoidal oscillatory environments follow $\gamma(t)=\gamma_0+A_\gamma \sin{(2\pi t / T_\gamma)}$, square wave oscillations follow $\gamma(t)= \gamma_0+A_\gamma\operatorname {sgn} \left(\sin {(2\pi t /T_\gamma)}\right)$, and randomly fluctuating environments follow the continuous uniform distribution $\gamma(t)=\displaystyle U\{\gamma_0-A_\gamma,\gamma_0+A_\gamma \}$, where $\gamma=\{\alpha,\beta\}$. 
When one parameter is constant and the other oscillates (Fig.~\ref{fig3}A,B), $\beta$ becomes the most limiting factor. If $\beta$ remains constant over time, only insignificant fractions of the original sequence O are replicated as R polymers. In contrast, a wide range of constant $\alpha$ values is compatible with the production of long C and R ssRNA molecules and allows the replication of the entire initial molecule O (Fig.~\ref{fig3}A). 

Fig.~\ref{fig3}C and D show how simultaneous (but not necessarily similar) fluctuations in both $\alpha$ and $\beta$ parameters affect the formation and length of C and R polymers. Results in Fig.~\ref{fig3}C indicate that, for the different cases studied, maintaining a standard sinusoidal $\beta$ oscillation while varying the period and amplitude of sinusoidal $\alpha$ oscillations (as well as changing it to a more abrupt square wave) does not significantly affect replication efficiency. The mean length of the obtained R polymers remains close to a 50\% of the original O sequence, and the maximum value after 100 realizations of all oscillation patterns recovered the whole original sequence length (20 nts). 
In contrast, maintaining an $\alpha$ oscillation compatible with high polymerization activity in comp.I and varying the period or amplitude of $\beta$ oscillations drastically affects the system performance (Fig.~\ref{fig3}D), presenting from a strong enhancement to a collapse of the replicative dynamics. 
In summary, EarlyWorld shows that efficient template replication of RNA is only plausible under oscillating environments, being their proficiency much more dependent on the properties of the nt-nt interactions that rule the template replication than on the clay-nt interplay that controls the initial RNA polymerization.

\begin{figure}[ht!]
\centering
\includegraphics[width=\linewidth]{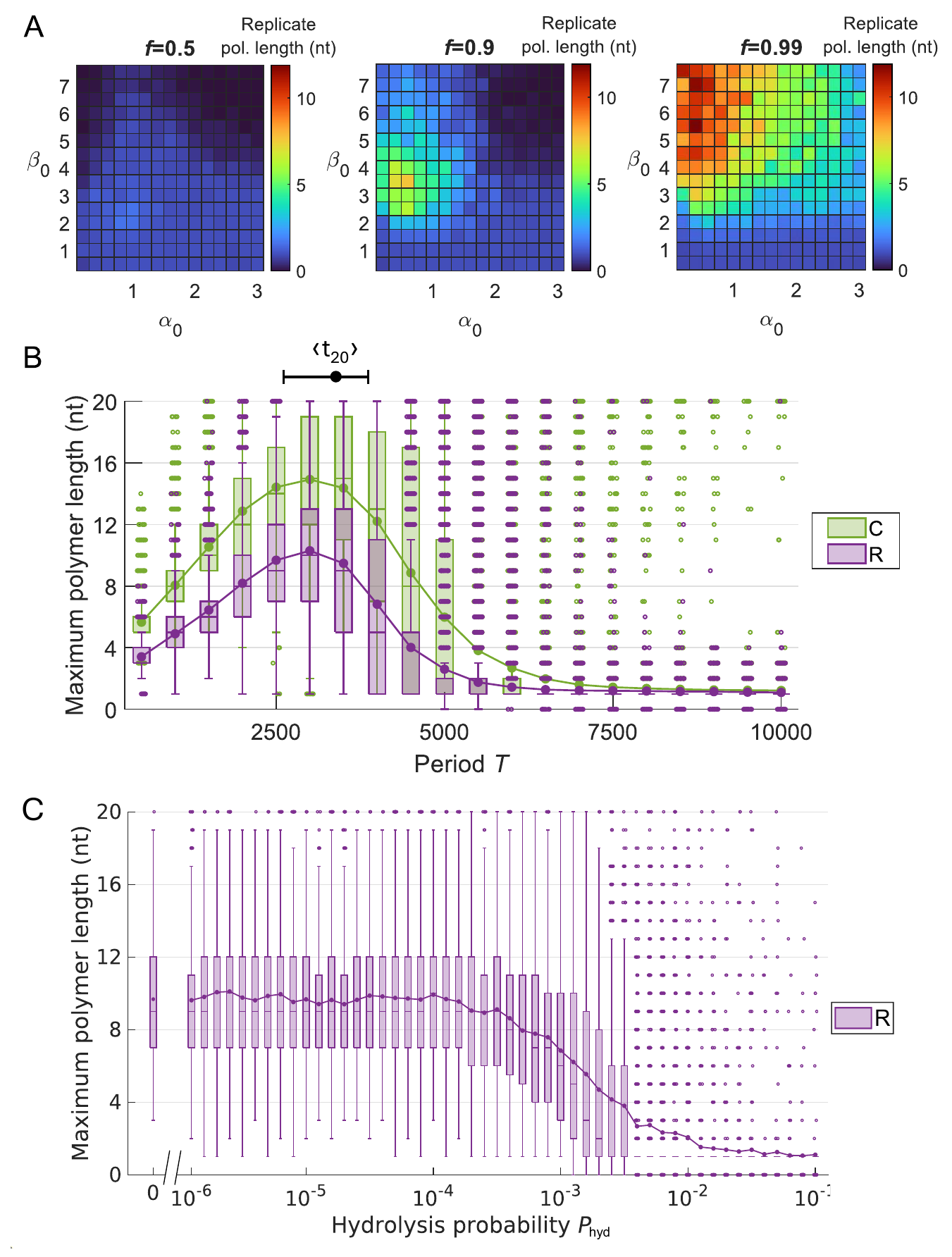}
\caption{Effect of the model parameters on the template-dependent replication of a ssRNA molecule adsorbed to a clay surface (comp.II). (A) Dependence of the mean of maximum R polymer lengths with average interaction parameters $\alpha_0$ and $\beta_0$ at three different amplitude fractions $f$ (being $A_{\alpha}=f\alpha_0$ and $A_{\beta}=f\beta_0$). $T_{\alpha}=T_{\beta}=2500$. (B) Dependence of the maximum length distribution of C polymers (complementary, green) and R polymers (replicates, purple) on the oscillation period $T=T_\alpha=T_\beta$. $\alpha_0=0.8$, $A_{\alpha}=0.5$ and $\beta_0=6$, $A_{\beta}=5.9$. The mean time to polymerize a 20-nt long ssRNA by template-dependent polymerization $\langle t_{20} \rangle$ is displayed with 25th-75th percentiles (see Supplementary Fig.~S6). (C) Dependence of the maximum length distribution of R polymers on the phosphodiester bond hydrolysis probability per time step $P_\mathrm{hyd}$. 
$\alpha_0=0.8$, $A_{\alpha}=0.5$ and $\beta_0=6$, $A_{\beta}=5.9$, $T=T_\alpha=T_\beta=2500$. Every simulation started with a random 20 nt-long ssRNA molecule, and was repeated 20 times in (A), $10^4$ times in (B) and 500 times in (C)}.
\label{fig4}
\end{figure}

In Fig.~\ref{fig4}, a more complete scan of the space of parameters for the template-dependent RNA replication that takes place in comp.II is presented. 
Without loss of generality, it is now assumed that (i) in natural environments the whole system formed of comp.I and II of a clay grain in contact with water will be exposed to the same oscillatory environmental conditions, thus affecting interaction parameters $\alpha$ and $\beta$ with the same period $T_\alpha=T_\beta=T$, and (ii) the amplitude of such oscillations will be proportional to the average values $\alpha_0$ and $\beta_0$ by a factor $f\in[0,1]$ (i.e., $A_\alpha=f\alpha_0$ and $A_\beta=f\beta_0$). Figure~\ref{fig4}A shows the dependence of the replication efficiency on the average interaction parameters $\alpha_0$ and $\beta_0$ and the amplitude fraction $f$ (see Supplementary Fig.~S5 for the corresponding plots on the efficiency of the formation of the complementary strand C). As already observed in Fig.~\ref{fig3}A, relatively low $\alpha_0$ is optimal because, in order to form Rs, there must be free space in the clay for Cs to adsorb and serve as Rs' templates, a condition better satisfied with an environment that hampers the attachment to the clay. On the other hand, large $\beta_0$ and very large amplitudes of the oscillations become necessary, as they represent environments that alternate periods of strong polymerization enhancement with periods of intense dsRNA denaturation. 

Fig.~\ref{fig4}B shows the dependence of the replication efficiency on the $\alpha$ and $\beta$ oscillation period. Short periods prevent from a sufficient elongation of the polymerized or replicated strands, the lower limit $T\rightarrow 0$ being comparable to a random distribution of $\beta$ values (as shown in Fig.~\ref{fig3}D). Intermediate period values favor the replicative dynamics, and copies of all lengths are densely represented. Remarkably, the value of the oscillation period that optimizes the replication dynamics ($3000\pm500$ time steps) in Fig.~\ref{fig4}B coincides with the number of simulation time steps that in average is required to form a complete 20-nt long complementary copy attached to the original O sequence ($\langle t_{20} \rangle\approx3300$, see Supplementary Fig.~S6). This equivalence will be of the maximum importance to match simulation times with real time scales and allow EarlyWorld to generate testable predictions about prebiotic environments (see Discussion). 
When oscillation periods become very large, a drastic transition occurs: the environment varies so slowly that the system resembles temporarily a $\beta$-constant environment (as shown in Fig.~\ref{fig3}B), favoring an excessive polymerization of dsRNA molecules that do not denature in such conditions, leading to the concomitant collapse of the replication process.
Interestingly, as it can be observed in Supplementary Fig.~S7, the maximum R polymer lengths observed for short periods follow log-normal distributions, become decreasing long-tailed distributions strongly peaked in 1 nt just after the transition, and finally collapse to the absence of any relevant replication for larger periods. 

Finally, Fig.~\ref{fig4}C shows the replication efficiency of the system when different values of the phosphodiester bond hydrolysis probability per time step, $P_\mathrm{hyd}$, are introduced in the model. It is known that 
the phosphodiester bond hydrolysis rate must be much lower than the polymerization rate to allow the formation of long polymers~\cite{Tkachenko2015}, and this ratio can vary widely depending on environmental conditions~\cite{Fernando2007,Bernhardt2012,Roy2020}. In particular, montmorillonite and other clays have been shown to reduce the rate of chemical hydrolysis of RNA~\cite{Ferris2006,aguilar2010radiation,zhang2023rna,saha2024effect}. Accordingly, we investigated a wide range of values for the ratio between the probability of hydrolysis $P_\mathrm{hyd}$ and the probability of phosphodiester bond formation $P_\mathrm{form}$ (which is equivalent to studying the dependence on $P_\mathrm{hyd}$ because in our simulations $P_\mathrm{form}=1$). When hydrolysis probability is below a critical threshold, the replication process proceeds as if hydrolysis was absent; however, around $P_\mathrm{hyd} \sim 10^{-4}$, the system begins to destabilize and replication becomes unsustainable beyond $P_\mathrm{hyd} \sim 10^{-3}$. 
A modest enhancement in replication efficiency is observed when the hydrolysis probability varies over time, consistent with the fluctuating conditions likely present in prebiotic environments (see Supplementary Fig.~S8).
Notably, the system's response to changes in both the oscillation period (Fig.~\ref{fig4}B) and the hydrolysis probability (Fig.~\ref{fig4}C) closely resembles the first-order transition previously described by Tkachenko and Maslov~\cite{Tkachenko2015}, which distinguished a regime sustained by long, self-replicating polymer populations from one dominated by free monomers depending on the overall monomer concentration. Furthermore, optimal periods in Fig.~\ref{fig4}B, together with the value of $\alpha_0$ used ($\alpha_0=0.8$), also optimized polymerization in comp.I (Fig.~\ref{fig2}). This strongly reinforces the plausibility of observing in real systems the mechanism here presented for the polymerization and posterior replication of RNA molecules. Also, Supplementary Fig.~S9 shows that the results presented in Fig.~\ref{fig4} and Supplementary Fig.~S6 remain robust under variations in either the available clay surface ---by changing the clay length $L$--- or in the nt concentration in the aqueous phase ---by modifying the number of molecules that fall into the clay per time step---.

\subsection{RNA replication across different genetic alphabets}
\label{sec:alphabets}

The computational environment EarlyWorld can be used to test multiple physico-chemical conditions compatible with those present in the early Earth where RNA world was established. 
One of the most controversial questions about the origin and evolution of our genetic material deals with the number, relative concentration and pairing rules of different monomers that formed RNA and, eventually, pre-RNA polymers (including glycerol-derived nucleic acid or GNA, threose nucleic acid or TNA, pyranosyl-RNA or p-RNA, locked nucleic acid or LNA, and peptide nucleic acid or PNA)~\cite{szathmary1991four,Szathmary2003, ruiz2014prebiotic, pinheiro2012synthetic}. Additionally, N-(2-aminoethyl)-glycine or AEG (building block of PNA backbone) could have been the precursor of current ribonucleosides, thus promoting the synthesis of proto-RNA~\cite{castanedo2024prebiotic}.

Current RNA is composed of four different nts, whose nucleobases interact specifically through hydrogen bonding: adenine and uracil (paired by means of two H bonds, A=U), as well as guanine and cytosine (linked by three H bonds, G$\equiv$C). In addition, non-canonical base pairs, such as Wobble base pairs (the most common type of these being G=U) are not infrequent in RNA, though they are less thermodynamically stable than canonical pairs.
In order to shed light on why prebiotic chemistry chose one genetic alphabet among all possible ones, we used our computational model to test the replication efficiency of six different options: two-letter alphabet [A=U] (A2$_\mathrm{au}$), two-letter [G$\equiv$C] (A2$_\mathrm{gc}$), canonical four-letter [A=U, G$\equiv$C] (A4), four-letter with additional Wobble base pair [A=U, G$\equiv$C, G=U] (A4$^*$, representing current RNA alphabet and pairing rules), and six-letter with a new base pair involving two unknown nucleotides (termed X and Y) linked by either two [A=U, G$\equiv$C, X=Y] (A$6_2$) or three hydrogen bonds [A=U, G$\equiv$C, X$\equiv$Y] (A$6_3$). 
The main properties of these alphabets when introduced in EarlyWorld are shown in Table~\ref{table1}.
\begin{table}[ht!]
\centering

\caption{Parameters of the computational environment EarlyWorld and the theoretical model for different genetic alphabets, from 2 to 6 letters. $s_\mathcal{A}$ and $t_\mathcal{A}$ are normalizations (such that $s_\mathcal{A}=t_\mathcal{A}=1$ for A4) of the average number of hydrogen bonds per nt and the connection time, respectively, used in the mathematical model.}
\begin{tabular*}{\hsize}{@{\extracolsep{\fill}}lcccc}
Alphabet & H-bonds/nt & $s_\mathcal{A}$ & Connection time & $t_\mathcal{A}$\cr
\hline
A2$_\mathrm{au}$&2 & 4/5&2&1/2\cr\hline
A2$_\mathrm{gc}$&3& 6/5&2& 1/2\cr\hline
A4&5/2& 1&4& 1\cr\hline
A4$^*$&7/3& 14/15&8/3& 2/3\cr\hline
A6$_2$&7/3& 14/15&6& 3/2\cr\hline
A6$_3$&8/3& 16/15&6& 3/2\cr\hline
\end{tabular*}
\label{table1}
\end{table}

Two factors play a role when comparing different genetic alphabets in EarlyWorld. The first is the connection time, that is, the average number of attempts (or time steps) that a nt takes to find a complementary nt to attach to, which is equal to the alphabet size $\mathcal{A}$ (with the exception of A4$^*$, which allows 3 kinds of base pairs with only 4 letters). The second one is the number of hydrogen bonds involved in each base pair: we now assume that the probability of denaturation of a dsRNA is proportional to the total number of hydrogen bonds formed between a template and its complementary strand, and not just to the length of the copied molecule, as considered before. Consequently, the denaturation probability of hybridized dsRNA (Eq.~\ref{Prup2}) becomes $P=e^{-n\beta/n_\mathrm{A4}}$, where $n$ represents the number of hydrogen bonds in the molecule and $n_\mathrm{A4}=2.5$ is the average number of bonds per nt in the A4 alphabet.

\begin{figure}[ht!]
\centering
\includegraphics[width=\linewidth]{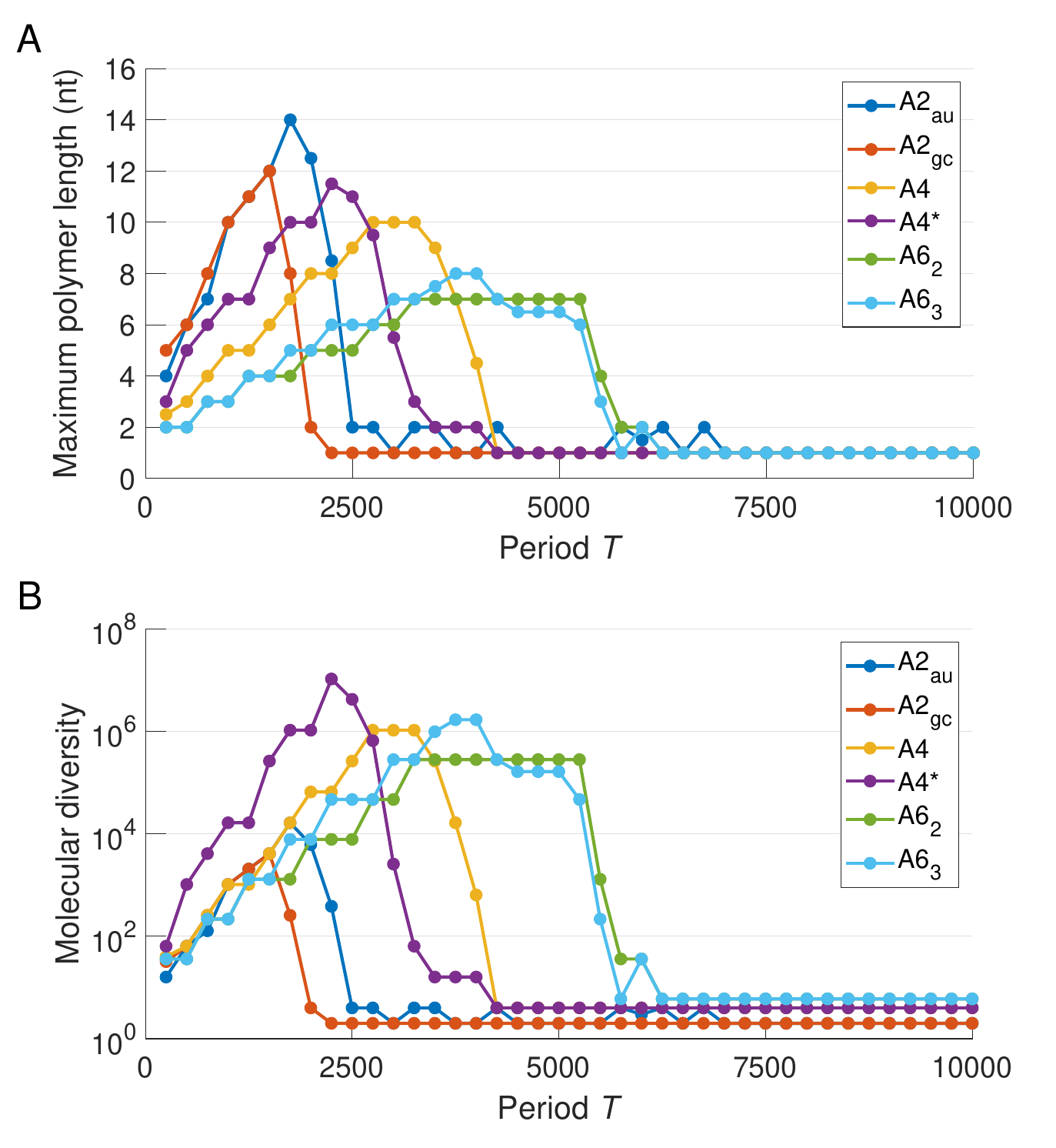}
\caption{Efficiency of template-dependent RNA replication is evaluated across six different alphabets. (A) Median of the distribution of maximum R polymer lengths, and (B) median of the molecular diversity computed as $\mathcal{A}^l$, where $\mathcal{A}$ stands for the alphabet size and $l$ is the median of the distribution of maximum R polymer lengths, for different values of the period $T=T_{\alpha}=T_{\beta}$. $\alpha_0=0.8$, $A_{\alpha}=0.5$ and $\beta_0=6$, $A_{\beta}=5.9$. Every simulation started with a random 20-nt RNA sequence, and was repeated 100 times. }
\label{fig5}
\end{figure}

In order to compare the replication efficiency of RNA composed of the 6 different alphabets under study, Fig.~\ref{fig5}A shows the dependence of the maximum length of the replicated strand R obtained by EarlyWorld with the oscillation period in an $\alpha-$ and $\beta-$sinusoidal environment (with system parameters equivalent to those in Fig.~\ref{fig4}B). 
The absolute maximum in replication length decreases with the alphabet size because of the slowing down of replicative dynamics (i.e., growth of connection time) for large alphabets. 
The effect related to the different number of hydrogen bonds between complementary nts makes G$\equiv$C-rich and X$\equiv$Y-rich dsRNA molecules more robust under denaturations and faster in polymerizing, thus accelerating the system dynamics and in consequence shifting the curves towards smaller periods (as it can be observed when comparing A2$_\mathrm{gc}$ vs. A2$_\mathrm{au}$ and A6$_3$ vs. A6$_2$).
Remarkably, A4$^*$ reaches significantly longer maximum polymer lengths than A4 because of its faster polymerization rate due to the formation of the non-canonical base pair G=U, performing in fact similar to A2$_\mathrm{gc}$. 

Additionally, if the total number of different ssRNA sequences that can be replicated with a length $l$ of $\mathcal{A}$ nts, $\mathcal{A}^l$, is used as a proxy for the molecular diversity ---or, if preferred, the information storage capacity--- associated with a population of RNA polymers of a certain genetic alphabet, Fig.~\ref{fig5}B shows that A4$^*$ represents an optimum combination in terms of diversity and replication speed, its absolute maximum outperforming the rest of alphabets in at least one order of magnitude. In turn, the absolute maximum molecular diversity associated with the canonical four-letter alphabet A4 is two orders of magnitude larger than that of two-letter alphabets, and similar to the much more energetically consuming six-letters ones.

\subsection{Theoretical approach to RNA replication in EarlyWorld}

We propose here a simplified theoretical description of the RNA polymerization and replication process where isolated nts can adsorb to, polymerize on and be released from the clay (level 0 in EarlyWorld, comp.I) or from a ssRNA already adsorbed to the clay (level 1 in EarlyWorld, comp.II). For simplicity, the model entails a unique interaction parameter $\gamma$ that plays the role of both $\alpha$ and $\beta$, depending on the case. It is assumed that the probability $P_i$ of forming and releasing a polymer of length $i$ to the aqueous phase or pool (level 2 in EarlyWorld) is equal to the probability of a single nt of remaining adsorbed to the surface (level 0 in comp.I and level 1 in comp.II) for one time step, receiving a new nt that lands adjacent to it (which we assume with probability 1 for simplicity), remaining the new 2-mer ssRNA adsorbed for one time step, and so on until the polymer reaches length $i$ after $i$ time steps and is released to the pool. In summary, this iterative calculation yields the probability $P_i$ of generating the sequences of length $i$ in a constant environment, as 
\begin{equation}
 P_1=e^{-\gamma};\quad P_i=e^{-i\gamma}\prod_{k = 1}^{i-1} (1-e^{-k\gamma}) \quad \mathrm{for}\,2\leq i\leq i_\mathrm{max},\label{Eq:Pi}
\end{equation}
where $i_\mathrm{max}$ should be large enough to make $P_{i_\mathrm{max}}$ negligible: $i_\mathrm{max}=50$ in all calculations. The mean length of the polymers produced for a given interaction parameter $\gamma$ is 
\begin{equation}
\bar{l}= \sum_{i=1}^{i_\mathrm{max}} i P_i\,, \label{Eq:lmedia}
\end{equation} 
a complex but totally explicit mathematical expression.
To model the clay-nt interaction in both comp.I and II in fluctuating environments, $\gamma=\alpha$ oscillates with time as 
\begin{equation}
\gamma=\gamma_0+A \sin{(2\pi i/ T)}\,,\label{Eq:osc} 
\end{equation} 
where $\gamma_0$ is the average interaction parameter, $A$ is the oscillation amplitude and $T$ the oscillation period. $T$ is an integer number between $T_\mathrm{min}=3$ (to avoid $\sin{(2\pi i/ T)}=0$ for all $i$) and $T_\mathrm{max}=i_\mathrm{max}/2$ (to make sure that the system reproduces at least two full oscillations). 
Note, however, that the nt-nt interactions in comp.II are influenced by the genetic alphabet in use due to complementary base pairing. In consequence, the evolution of $\gamma=\beta$ in fluctuating environments becomes 
\begin{equation}
\gamma=s_\mathcal{A} (\gamma_0+A \sin{(2\pi t_\mathcal{A} i / T))}\,, \label{Eq:alfabetos}  
\end{equation}
where $s_\mathcal{A}$ is proportional to the average number of hydrogen bonds between two nts for each alphabet, and $t_\mathcal{A}$ is proportional to the connection time that a nt in the pool would take in average to find its complementary one (see Table~\ref{table1}). 

\begin{figure}[ht!]
\centering
\includegraphics[width=\linewidth]{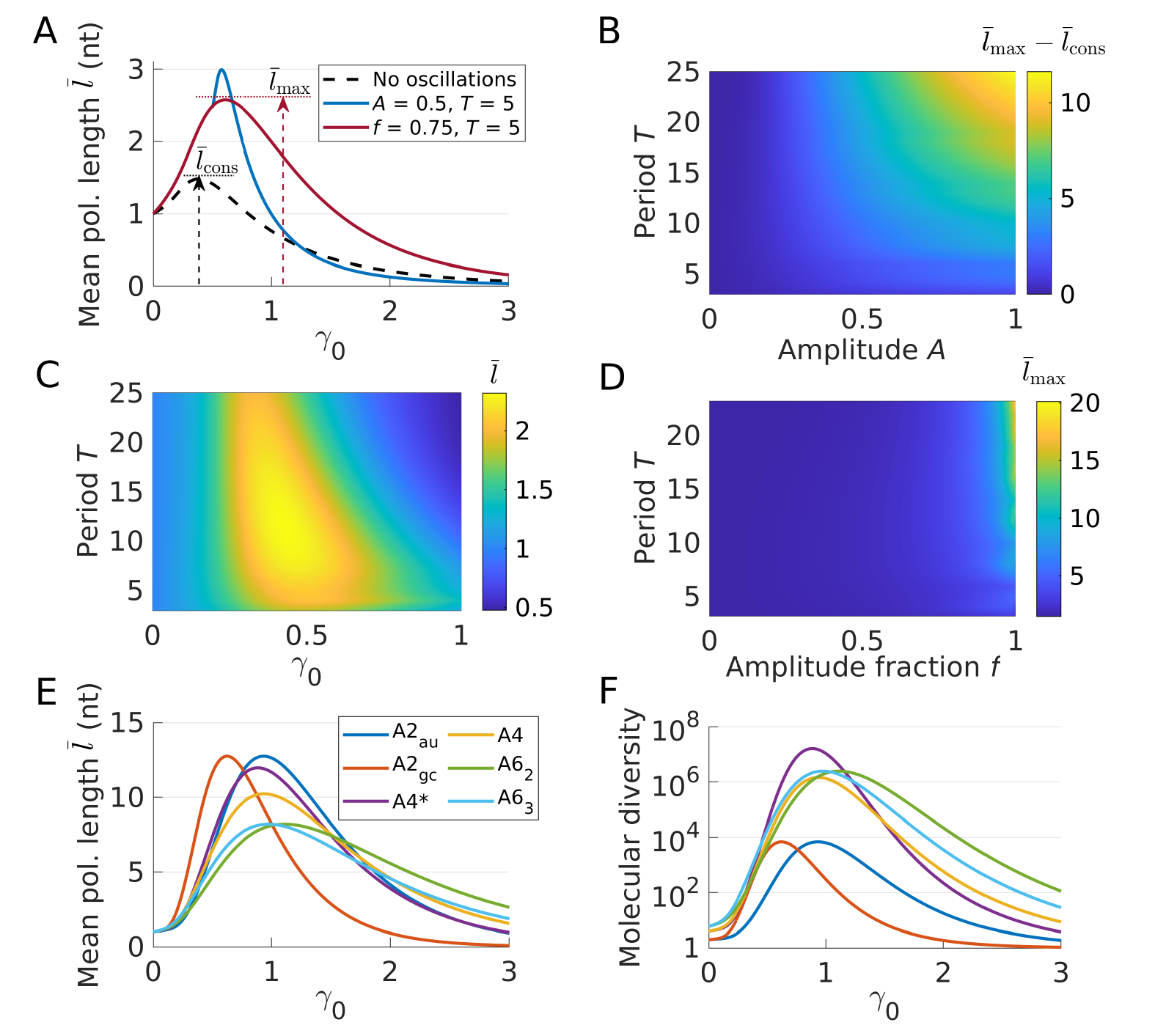}
\caption{Simplified mathematical approach to the polymerization and replication of a ssRNA molecule on clay surfaces. (A) Dependence of the mean polymer length $\bar{l}$ produced on the average interaction parameter $\gamma_0$ in three scenarios: a constant environment ($\gamma = \gamma_0$), an oscillating environment with amplitude $A=0.5$ ($\gamma_0 \geq A$ to avoid meaningless negative $\gamma$) and period $T=5$, and an oscillating environment of relative amplitude $f=0.75$ and $T=5$. The maximum mean length of the polymers produced in an oscillating environment $\bar{l}_{\text{max}}$ and in the constant environment $\bar{l}_{\text{cons}}=1.4823$ are remarked. (B) Difference between the maximum mean length produced in oscillating environments and in the constant environment for an exhaustive range of system parameters. Interestingly, all values of $\bar{l}_{\text{max}} - \bar{l}_{\text{cons}}$ are $> 0$. (C) Dependence of the mean length of polymers produced in oscillating environments $\bar{l}$ with the average interaction parameter $\gamma_0$ and the period $T$, $f=0.6$. (D) Dependence of the maximum mean length of the polymers $\bar{l}_{\text{max}}$ produced in oscillating environments with the amplitude fraction $f$ and the period $T$. (E) Mean length $\bar{l}$ of the polymers produced in oscillating environments for the six different genetic alphabets considered in EarlyWorld and a wide range of $\gamma_0$, $f=0.95$, and $T=25$. (F) Molecular diversity $\mathcal{A}^{\bar{l}}$ associated with the polymers produced for different genetic alphabets in (E).}
\label{fig6}
\end{figure}

Fig.~\ref{fig6}A plots the dependence of the mean length $\bar l$ of the polymers produced by the system on the average interaction parameter $\gamma_0$ (i.e., $\alpha_0$ in comp.I and II and $\beta_0$ in comp.II) for a constant environment (given by Eqs.~\ref{Eq:Pi}-\ref{Eq:lmedia} with $\gamma=\gamma_0$) and two different oscillating environments (given by Eqs.~\ref{Eq:Pi}-\ref{Eq:osc} for $\gamma_0=\alpha_0$ and \ref{Eq:Pi}-\ref{Eq:alfabetos} for $\gamma_0=\beta_0$).
All mean polymer lengths show a maximum ${\bar l}_\mathrm{max}$ for intermediate values of $\gamma_0$, as it happened in the EarlyWorld simulations shown in Figs.~\ref{fig2}B and \ref{fig3}A.
Note that the maximum of the mean lengths of both oscillating environments are larger than that of the constant one ${\bar l}_\mathrm{cons}=1.4823$. In order to verify theoretically whether this is a general behavior, Fig.~\ref{fig6}B shows the difference between the maximum of the mean lengths of the polymers produced in oscillating environments and constant environments (i.e., ${\bar l}_\mathrm{max}-\bar l_\mathrm{cons}$) for an exhaustive range of all the system parameters. This subtraction results to be always positive, proving that oscillatory environments are always more favorable than constant ones regarding the abiotic polymerization and replication of RNA sequences. 

Despite the simplicity of this mathematical model in comparison to the EarlyWorld computational environment, it captures several of its main findings, reinforcing its generality. 
As an example, Fig.~\ref{fig6}C shows the dependence of the mean length of the polymers produced by the system with the interaction parameter $\gamma_0$ and the oscillation period $T$, where it is clear that for any meaningful value of the environment (i.e., of the physico-chemical parameters that influence the clay-nt and nt-nt molecular interactions), the system shows the maximum replication efficiency for intermediate values of $T$, as it was shown in Fig.~\ref{fig4}B for EarlyWorld. Figure~\ref{fig6}D is focused on the influence of the amplitude fraction $f$, remarking that only strongly oscillatory environments give rise to significant replication dynamics, in agreement with Fig.~\ref{fig4}A for the numerical model. Finally, Fig.~\ref{fig6}E,F compare the proficiency of six genetic alphabets in the polymerization and replication of primordial RNA molecules. Here, the mathematical model recovers precisely the results obtained for EarlyWorld in Fig.~\ref{fig5}, as: (i) faster hybridization dynamics of small alphabets (composed of 2 or 4 nts) make them generically outperform large alphabets; and (ii) regarding molecular diversity, the absolute maximum of alphabet A4 coincides in performance with any larger alphabet and drastically outperforms 2-letter ones, while A4$^*$ shows an unbeatable combination of alphabet length and speed of replication dynamics. 

\section{Discussion} 
The transition from astrochemistry to prebiotic chemistry and then to life on Earth involved critical steps toward increasing molecular complexity, with the origin of genetic information and its template-dependent replication being a key, unresolved question~\cite{Szathmary1995, Garcia_2022}. Motivated by this challenge, we developed EarlyWorld, a computational model that simulates primordial RNA polymerization and replication, driven solely by the interactions at the clay-water interfaces. The model features two interconnected compartments within the clay, enabling the separation of polymerization and replication processes, which is difficult to achieve in other heterogeneous, plausibly prebiotic environments such as lipid-water or air-water interfaces, and ice-water eutectic phases~\cite{ruiz2014prebiotic, ianeselli2023physical}. 
The system's performance depends on two parameters that represent the strength of clay-nt ($\alpha$) and nt-nt ($\beta$) interactions. Given that fluctuating, out-of-equilibrium conditions likely played a key role in life's origins~\cite{ianeselli2023physical}, we explored a range of oscillatory patterns by varying $\alpha$ and $\beta$ both periodically and aperiodically. 

Despite random elongation of some RNA sequences~\cite{Fernando2007, Chamanian2022}, our results show that 20-mer RNA oligonucleotides can polymerize and replicate with a 100\% accuracy under moderate phosphodiester bond hydrolysis rates (Fig.~\ref{fig4}C) and specific oscillating environmental conditions (Supplementary Fig.~S3), whose compatibility across both compartments of the clay system suggests that all steps of polymerization and replication could occur within a single physico-chemical environment. In particular, temperature, pH, salinity, or humidity should fluctuate with (i) large amplitudes (Fig.~\ref{fig4}A), and (ii) periods matching the time necessary to polymerize a complementary ssRNA of the same length as the template (Fig.~\ref{fig4}B, Supplementary Fig.~S6), being the results compatible with a diversity of nucleotide concentrations in solution and available clay surfaces (Supplementary Fig.~S9). Furthermore, there is still an open debate on whether RNA polymerization on different clay substrates, as described by the groups of Ferris and others, preferentially occurs within the phyllosilicate interlayers or on the more accessible external surfaces~\cite{ferris1996synthesis, huang2003synthesis,franchi2005surface,huang2006one, Ferris2006,joshi2009mechanism,kaddour2018nonenzymatic,dujardin2022formation}. The EarlyWorld model produces similarly efficient polymerization and replication if the polymerization scenario corresponds to a sterically unrestricted surface (e.g., a wide internal channel, inter-particle site or external clay surface, see Supplementary Fig.~S10A and B), provided that it is sufficiently separated from the replication site. In contrast, when polymerization and replication occur at the same location on the clay surface, competition among RNA polymers for limited surface sites significantly reduces replication efficiency (Supplementary Fig.~S10C).

The ability of our model to generate testable predictions about environments that could optimize primordial RNA replication hinges on result (ii) above, combined with recent geochemical insights into Hadean Earth conditions~\cite{Farhat:2022} and the estimates of template-dependent RNA polymerization times at clay-water interfaces. Our results yield that the optimal environmental oscillation period for the prebiotic template-dependent replication of a ssRNA of length $l$ would be that which coincides with the time to obtain a polymer complementary to the original one. Therefore, if we obtained an experimental estimate of 
the template-dependent RNA polymerization time per nt in a clay-water environment $t_\mathrm{pol}$, the ideal prebiotic environment for the polymerization and replication of a ssRNA of length $l$ would be that with an oscillation period equal to
\begin{equation}
    T_\mathrm{opt}=l\times t_\mathrm{pol}\,,
    \label{Eq:Topt}
\end{equation}
assuming a linear relationship between RNA polymerization time and sequence length. Though precise experimental data for such polymerization time per nt remain yet unavailable, relevant experiments offer insight. Extensive work by Ferris and collaborators~\cite{ferris1996synthesis, huang2003synthesis,huang2006one,Ferris2006} demonstrates that RNA polymerization time on montmorillonite clay varies significantly, from 0.6 to 11 hours per nt, depending on nt types, concentrations and presence of different activating leaving groups linked to the phosphates ~\cite{huang2006one}.
Additionally, nonenzymatic template-dependent RNA polymerization in aqueous solution without mineral substrates yields a polymerization time of approximately 1 h/nt~\cite{wu1992nonenzymic,blain2014progress,li2017enhanced}, which may extend to 1 day/nt under conditions that increase their prebiotic plausibility~\cite{ding2023experimental}. Furthermore, we can consider a hydrolysis time of $\sim20-50$ h/nt~\cite{blain2014progress}, which helps us to fix an upper bound for $t_\mathrm{pol}$ because polymerization times larger than the hydrolysis time would not permit a neat sequence polymerization. Taking into account these data, we can expect an overall clay-catalyzed template-dependent RNA polymerization time in prebiotic conditions of $t_\mathrm{pol}$ in the range $\sim1$ h/nt $-$ 1 day/nt. 
In summary, based on the available experimental data and Eq.~\ref{Eq:Topt}, we propose that plausible oscillation periods would range from around one current day (i.e., 24h-long) to a few weeks to support efficient clay-catalyzed replication of ssRNA oligomers long enough to be functional, though short enough to be polymerized and replicated in prebiotic conditions (i.e., $\sim15-50$ nt) at the dawn of the RNA world.

Recent and precise analyses of tidal interactions in the Earth-Moon system since its formation~\cite{Farhat:2022} suggest that the planet-satellite distance was 50-62\% of its current value (Supplementary Fig.~S11A) and that a day lasted about 10-12 h when life emerged on Earth, approximately 4.2-3.8 Gyr ago~\cite{Moody:2024,kitadai2018origins,pearce2018constraining} (Supplementary Fig.~S11B). The optimal oscillation periods for RNA replication, as discussed earlier, would probably exceed the timescales of daily temperature fluctuations from early Earth's day-night cycles or tidal wet-dry dynamics but would be shorter than seasonal variations in temperature and humidity. 
These oscillations, however, align well with spring tides: the maximal tides that occur twice each synodic month (the period between two Sun-Earth-Moon alignments), with a spring tide period of around 125-175 h when life emerged (Supplementary Fig.~S11C). Unlike other potential fluctuating prebiotic scenarios, tidal pools benefit from the strictly periodical behavior of tidal forces, and have already been considered potentially relevant for prebiotic chemistry~\cite{lathe2004fast,lathe2005tidal}. These pools would have facilitated interactions between a wide array of reactants carried by rivers, oceans, or the atmosphere~\cite{kitadai2018origins}, on a planet where isotopic evidence from detrital zircons indicates that continental crust may have formed as early as 4.35 Gyr ago~\cite{Harrison:2008, Stueken:2013}. Given that ---to first order--- tidal amplitudes decrease with the cube of the Earth-Moon and Earth-Sun distances~\cite{macdonald1964tidal}, the Moon's contribution was 4 to 8 times greater than today, while the Sun's contribution has remained constant, as the Earth-Sun distance has not substantially changed since then. 

Therefore, in the model we envisage, every 10-18 early Earth's days warm seawater would flood dry clay-bottomed ponds covered by salt precipitates, located far enough from the shoreline to be unaffected by daily tides. This seawater influx could have three key effects: (i) delivering fresh nts from the sea or nearby sources; (ii) increasing the temperature at clay-water interfaces; and (iii) reducing the local concentration of monovalent and divalent cations that were present in the salt precipitates. Together, (ii) and (iii) increased the medium's astringency (i.e., reduced $\alpha$ and $\beta$ in the EarlyWorld model), weakening nt hydrogen bonding and promoting RNA strand separation. These findings are consistent with previous results~\cite{ianeselli2023physical}, where temperatures above $60^{\circ}$C, Na$^+$ concentrations below 100 mM, and Mg$^{2+}$ concentrations below 5 mM promote dsRNA denaturation. 
After several days of no further seawater input, the (perhaps, warm little) pond would dry out again (i.e., increasing $\alpha$ and $\beta$ in the EarlyWorld model), allowing nts to polymerize on the clay surfaces as well as on the adsorbed ssRNA templates, thus contributing to the overall template-dependent replication. Thanks to the periodic behavior of spring tides, this cyclic process of RNA polymerization and replication could continue without interruption. Supplementary Fig.~S11D presents our model's prediction for the RNA sequence lengths that would be optimally replicated in a spring tidal pool, based on the polymerization time (a yet very uncertain quantity) and the Earth's time. The range 4-11 h/nt for polymerization times matches the 15-40 nt lengths typically obtained in EarlyWorld. These results encourage experimental efforts to determine precise template-dependent RNA polymerization times in clay-containing prebiotic environments, as they would provide relevant insight into the expected length of primordial RNA oligomers.

In addition, our numerical and theoretical results reveal that the four-letter alphabet with pairing rules [A=U, G$\equiv$C, G=U] (A4$^*$, representing current RNA) exhibits superior replication performance than any other alphabet in terms of molecular diversity and sequence space navigability. The canonical four-letter alphabet [A=U, G$\equiv$C] (A4) achieves molecular diversity two orders of magnitude greater than two-letter alphabets, and comparable to the more energetically costly six-letter systems. This strongly supports the primordial selection of four-letter alphabets (especially A4$^*$) for genetic information storage in RNA. Interestingly, A4$^*$ introduces sequence mutations through base pair degeneracy 
(e.g., an A copied as U in the complementary polymer may become G in the replicated one)
that expands dramatically  the sequence exploration ability, while remaining constrained by the entropy induced by the degeneration bias (see Supplementary Note S4 for an analysis of the expansion of sequence space exploration enabled by non-canonical base pairs through the lens of information theory). Note that, while this mutation potential could enhance molecular diversity in early RNA populations, it also risks an error catastrophe (according to quasispecies theory~\cite{eigen1978hypercycle}) in case non-canonical base pairs became too prevalent. Actually, G=U interactions are weaker than canonical ones, which may have helped to mitigate this risk. 

The precise copies of the original ssRNA produced through the process analyzed in this work should undergo subsequent selection pressures, which recent findings suggest could occur even for 22-nt long oligonucleotides~\cite{bartolucci2023sequence}. The study of such selection on the basis of the potential functionality of their molecular structures becomes a promising direction for future research. This will be addressed in the next version of EarlyWorld, thanks to the inclusion of an {\it in silico} RNA 2D-folding routine~\cite{ViennaPackage:2011}. Additional enhancements, such as varying the relative initial nt abundances, differential clay adsorption of purine and pyrimidine nts, allowance for non-productive $2'$-$5'$ and $5'$-$5'$ phosphodiester bonds (besides the regioselective formation of $3'$-$5'$ ones), and the introduction of replication errors (mismatches, insertions, deletions, and recombination), will help to bring the computational model closer to the rules of real (bio)chemistry. 

Recent studies have modeled similar phenomenologies using chemical kinetics, a perspective that aligns with our effective approach. In our framework, the desorption and denaturation probabilities, $P_{\alpha}$ and $P_{\beta}$, serve as simplified proxies for the kinetic constants of RNA desorption from the clay and dsRNA denaturation, respectively. For instance, the kinetic constant for RNA strand denaturation $k_\mathrm{off}$ is commonly assumed to depend on the binding free energy per base pair, $\Delta{G}$, as $k_\mathrm{off} \propto \exp(-\frac{\Delta{G}}{k_\mathrm{B}T}l)$, where $k_\mathrm{B}$ is Boltzmann’s constant, $T$ is the temperature, and $l$ is the number of base pairs in the RNA duplex~\cite{Chamanian2022,Goppel2022}. In our model, denaturation probability is $P_{\beta} = \exp(-\beta l)$, establishing an explicit connection with $k_\mathrm{off}$, where the parameter $\beta$ encapsulates the thermodynamic contribution to strand separation. 
A similar thermodynamic interpretation applies to $\alpha$, which depends on the binding free energy of nucleotide–clay interactions governing desorption. These simplifications enable a tractable model that incorporates environmental variability in a thermodynamically consistent manner, while inevitably omitting chemical details which are beyond the scope of this study.

Finally, we have developed a theoretical study of a toy model of the system that, despite its simplicity, captures most of the key phenomenological features observed in the computational framework, suggesting the generality of our results beyond specific numerical details. The model highlights the importance of intermediate oscillatory periods and strong fluctuations in nt-nt hydrogen bonding to facilitate RNA replication, as well as the dominance of four-letter alphabets in a potential primordial soup analogous to that envisioned by A.I. Oparin a century ago. Notably, the model provides strong evidence that RNA polymerization and replication are enhanced in oscillating environments compared to constant conditions at aqueous-clay interfaces. The Moon’s role in providing such oscillations in early Earth may have been crucial for the emergence of heritable information, and suggests that exoplanets with surface liquid water in interaction with clays (or, eventually, other rocks that provide catalytic surfaces) and orbited by large moons could be promising targets in the search for extraterrestrial life.

\section*{Data availability} 
The authors declare that the relevant data supporting the findings of this study are available within the paper and its supplementary information files. All data are available from the corresponding author upon reasonable request.

\section*{Code availability} 
An implementation of the computational framework EarlyWorld developed in Matlab(R2020b), with description, documentation and instructions of use is available at the following public open source repository: \url{https://github.com/c-alejandrev/EarlyWorld}. 

\section*{Author Contributions} 
C.A, A.A, C.B., and J.A. designed the study, conceived the computational environment EarlyWorld, discussed the results and wrote the paper; C.A, A.A and J.A. implemented the algorithm; C.A and A.A performed the numerical experiments; J.A. and A.A. developed the analytical work; and C.B. explored the connections with experimental prebiotic chemistry. The authors declare no conflict of interest.

\begin{acknowledgments}
The authors are indebted to A. de la Escosura, M. Fern\'andez-Ruz, and R. Guantes for critical reading of the manuscript;  to B. Corominas-Murtra for valuable comments and support on information theory; and to J. Aguirre-Becerril, M. Castro, P. Catal\'an, J.A. Cuesta, J. Iranzo, S. Manrubia, S. Rasmussen and M. Villani for their useful comments. J.A. and C.A. received support from grant No. PID2021-122936NB-I00, C.B. from grant No. PID2022-139908OB-I00, and A.A. from grant No. MDM-2017-0737 Unidad de Excelencia ``Mar\'{i}a de Maeztu'' - Centro de Astrobiolog\'{i}a (CSIC-INTA), all of them funded by the Spanish Ministry of Science and Innovation/State Agency of Research MCIN/AEI/10.13039/501100011033 and by ‘‘ERDF A way of making Europe’’. J.A. acknowledges financial support from grant No. PIE2024ICT085 funded by the Spanish National Research Council (CSIC), C.A. acknowledges the support of the Consejería de Educación, Ciencia y Universidades de la Comunidad de Madrid through grants No. PEJ-2021-AI/TIC-22450 and PIPF-2023/TEC29607 and A.A. the support of the field of excellence `Complexity of life in basic research and innovation' of the University of Graz. Authors benefited from the interdisciplinary framework provided by CSIC through the `LifeHUB.CSIC' initiative (PIE 202120E047-Conexiones-Life).
\end{acknowledgments}


%

\end{document}